\documentclass[aps,prb,reprint,superscriptaddress,floatfix,showpacs]{revtex4-1}

\usepackage[dvipdfmx]{graphicx}
\usepackage{bm}
\usepackage{color}
\usepackage{amsmath,amssymb}
\usepackage{array}

\renewcommand{\k}{{\bm k}}

\begin{document}

\title{Nonsymmorphic Weyl superconductivity in UPt$_3$ based on E$_{2u}$ representation}

\author{Youichi Yanase}
\email[]{yanase@scphys.kyoto-u.ac.jp}
\affiliation{Department of Physics, Graduate School of Science, Kyoto University, Kyoto 606-8502, Japan}

\date{\today}

\begin{abstract}
We show that a heavy fermion superconductor UPt$_{3}$ is a topological Weyl superconductor with tunable Weyl nodes. 
Adopting a generic order parameter in the $E_{\rm 2u}$ representation allowed by nonsymmorphic crystal symmetry, 
we clarify unusual gap structure and associated topological properties. 
The pair creation, pair annihilation, and coalescence of Weyl nodes are demonstrated 
in the time-reversal symmetry broken B-phase.  
At most 98 point nodes compatible with Blount's theorem give rise to line node-like behaviors in low-energy excitations, 
consistent with experimental results. 
We also show an {\it arc} node protected by the nonsymmorphic crystal symmetry on the Brillouin zone face. 
\end{abstract}

\maketitle

Superconductivity with nontrivial symmetry and topology is attracting renewed interest stimulated by 
enormous studies of topological insulators and superconductors~\cite{Kane-Mele,Schnyder,Kitaev2009,Qi-Zhang}. 
Strongly correlated electron systems are platform of such unconventional superconductivity~\cite{Yanase_review}. 
A heavy fermion superconductor UPt$_3$ discovered in 1980's~\cite{Stewart} unambiguously exhibits exotic properties, 
that is, multiple superconducting phases in the field-temperature plane [Fig.~1(a)]~\cite{Fisher,Bruls,Adenwalla}. 
The presence of the multiple superconducting phases is a direct evidence for a multi-component non-$s$-wave 
order parameter~\cite{Sigrist-Ueda}. 
Comparison between experiments and theories points to odd-parity 
spin-triplet superconductivity~\cite{Sauls,Joynt,Tou_UPt3}. 
Because odd-parity superconductivity is often accompanied 
by topological order~\cite{Schnyder,Kitaev2001,Sato2010,Tanaka_review}, 
it may be interesting to clarify the topological properties of UPt$_3$.

\begin{figure}[htbp]
\begin{center}
\includegraphics[width=80mm]{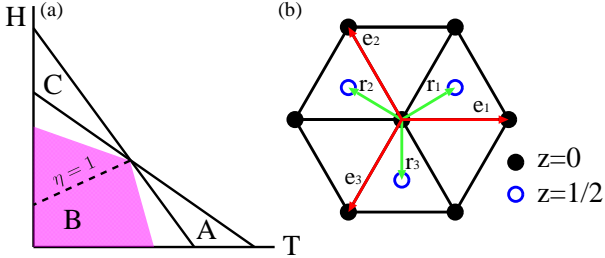}
\caption{(Color online)
(a) Multiple superconducting phases in UPt$_3$. Shaded region shows the Weyl superconducting phase. 
(b) Crystal structure of UPt$_3$. 
Uranium ions form AB-stacked triangular lattice. 
Two-dimensional vectors, ${\bm e}_i$ and ${\bm r}_i$, are shown by arrows. 
} 
\label{schematic}
\end{center}
\end{figure}

Topological order in superconductors is closely related to the symmetry and nodal structure of superconductivity. 
Despite intensive studies for three decades, the symmetry of superconductivity in UPt$_3$ is still under debate. 
However, a chiral $f$-wave state with orbital angular momentum $L_z = \pm 2$~\cite{Sauls} 
allowed in hexagonal crystals [see Fig.~1(b)] 
is supported by nodal excitations~\cite{Joynt} and broken time-reversal symmetry~\cite{Schemm} as well as 
by a phase sensitive measurement~\cite{Strand}. 
On the other hand, a recent thermal conductivity measurement points to   
another $f$-wave state with $L_z = \pm 1$~\cite{Machida-Izawa}. 

In spite of these intensive studies, the $f$-wave pairing states are incompatible 
with the Blount's theorem~\cite{Blount,Kobayashi-Sato} 
which proves the absence of line node in odd-parity superconductors~\cite{Comment1}. 
Although line node behaviors have been observed in UPt$_3$~\cite{Joynt},  
Blount's theorem implies that line nodes are fragile 
against perturbation preserving the symmetry of the system. 
Indeed, the line nodes disappear as a result of the mixing of $f$-wave Cooper pairs with $p$-wave ones 
in the same irreducible representation. 
According to the symmetry classification~\cite{Sigrist-Ueda}, not only the chiral $f$-wave state 
with $L_z = \pm 2$~\cite{Sauls} but also a $p$-wave state (and the $f$-wave state with 
$L_z = \pm 1$~\cite{Machida-Izawa}) belong to the $E_{\rm 2u}$ representation of $D_{\rm 6h}$ point group. 
Therefore a generic $E_{\rm 2u}$ state is induced by a mixed $p$+$f$-wave Cooper paring. 
In this paper we clarify the nodal gap structure  and specify the topological properties of a generic $E_{\rm 2u}$-state in UPt$_3$.  
It is revealed that the B-phase is a Weyl superconducting state~\cite{Meng} 
analogous to Weyl semimetals~\cite{Murakami,Wan-Vishwanath,Burkov-Balents} discovered recently~\cite{Xu,Lv,Yang,Huang}. 
We furthermore demonstrate pair creation, pair annihilation, and coalescence of Weyl nodes which do not occur in other chiral Weyl 
superconductors~\cite{Volovik,Sau-Tewari,Fischer_Weyl,Goswami}.

Our study is based on a Bogoliubov-de Gennes (BdG) Hamiltonian, 
\begin{eqnarray}
{\cal H}_{\rm BdG}&=& \sum_{{\bm k},m,s} \xi({\bm k}) c_{{\bm k}ms}^\dagger c_{{\bm k}ms} 
+ \sum_{{\bm k},s} \left[a({\bm k}) c^\dagger_{{\bm k}1s}c_{{\bm k}2s} + {\rm h.c.}\right]
\nonumber \\
&& \hspace{-5mm} +\sum_{{\bm k},m,s,s'} \alpha_m {\bm g}({\bm k}) \cdot {\bm s}_{ss'}c^\dagger_{{\bm k}ms}c_{{\bm k}ms'}
\nonumber \\ 
&& \hspace{-5mm} + \frac{1}{2} \sum_{{\bm k},m,m',s,s'} \left[\Delta_{mm'ss'}({\bm k}) c^\dagger_{{\bm k}ms}c^\dagger_{-{\bm k}m's'} + {\rm h.c.}  \right], 
\label{eq:model}
\end{eqnarray}
where ${\bm k}$, $m=1,2$, and $s=\uparrow,\downarrow$ are momentum, sublattice, and spin, respectively. 
Based on the crystal structure of UPt$_3$ illustrated in Fig.~\ref{schematic}(b), 
we adopt an intra-sublattice kinetic energy, 
$
\xi({\bm k}) = 2 t \sum_{i=1,2,3}\cos{\bm k}_\parallel\cdot{\bm e}_i + 2 t_z \cos k_z -\mu, 
$
and inter-sublattice hopping term,  
$
a({\bm k}) = 2 t' \cos\frac{k_z}{2} \sum_{i=1,2,3} e^{i{\bm k}_\parallel\cdot{\bm r}_i}, 
$
with ${\bm k}_\parallel=(k_x,k_y)$ and ${\bm e}_1 = (1,0)$, ${\bm e}_2 = (-\frac{1}{2},\frac{\sqrt{3}}{2})$, 
${\bm e}_3 = (-\frac{1}{2},-\frac{\sqrt{3}}{2})$, ${\bm r}_1 = (\frac{1}{2},\frac{1}{2\sqrt{3}})$, 
${\bm r}_2 = (-\frac{1}{2},\frac{1}{2\sqrt{3}})$, and ${\bm r}_3 = (0,-\frac{1}{\sqrt{3}})$. 
Since $D_{\rm 3h}$ local symmetry at Uranium ions lacks inversion symmetry, Kane-Mele spin-orbit coupling (SOC) with 
${\bm g}({\bm k})= \hat{z} \sum_{i=1,2,3} \sin{\bm k}_\parallel\cdot{\bm e}_i$~\cite{Saito_MoS2} 
appears in a sublattice-dependent way. 
The coupling constant is $(\alpha_1,\alpha_2)=(\alpha,-\alpha)$ so as to preserve the global $D_{\rm 6h}$ symmetry~\cite{Kane-Mele,Fischer,JPSJ.81.034702}. 
The inter-sublattice SOC is prohibited because the inter-sublattice bonds respect the inversion symmetry.

Quantum oscillation measurements combined with band structure 
calculations~\cite{Joynt,Taillefer1988,Kimura_UPt3,McMullan,Nomoto} have shown a pair of FSs 
centered at the $A$-point ($A$-FSs), three FSs at the $\Gamma$ point ($\Gamma$-FSs), 
and two FSs at the $K$ point in UPt$_3$. 
Since small FSs enclosing the $K$ point give a small density-of-states (DOS), 
they may play a minor role. 
Therefore, we study the superconducting properties of $A$-FSs and $\Gamma$-FSs one by one.
By choosing a parameter set $(t,t_z,t',\alpha,\mu)=(1,-4,1,2,12)$ our two band model reproduces 
a pair of $A$-FSs, while another set $(t,t_z,t',\alpha,\mu)=(1,4,1,0,16)$ reproduces 
topology of $\Gamma$-FS. Since the SOC is negligible for the $\Gamma$-FS, 
we simply set $\alpha=0$ in the latter parameter set.

Order parameter of the $E_{\rm 2u}$-state is generally represented by  
$
\hat{\Delta}({\bm k}) = \eta_1 \hat{\Gamma}_1 + \eta_2 \hat{\Gamma}_2 
$
with basis functions $\hat{\Gamma}_1$ and $\hat{\Gamma}_2$ composed of several components. 
Although the purely $f$-wave state has been intensively investigated~\cite{Sauls}, 
an admixture of a $p$-wave component is allowed by symmetry. 
Besides these components, a sublattice-singlet spin-triplet $d$-wave component naturally accompanies 
the $f$-wave component because of the nonsymmorphic crystal structure of UPt$_3$~\cite{Supplemental1}. 
Taking into account all the components, we study the $E_{\rm 2u}$-state with 
\begin{eqnarray}
&&
\hat{\Gamma}_1 = \bigl[\delta \left\{p_x({\bm k})s_x - p_y({\bm k})s_y\right\} \sigma_0 
\nonumber \\ && \hspace{10mm} 
+ f_{(x^2-y^2)z}({\bm k})s_z \sigma_x - d_{yz}({\bm k})s_z \sigma_y \bigr] i s_y,
\\
&&
\hat{\Gamma}_2 = \bigl[\delta \left\{p_y({\bm k})s_x + p_x({\bm k})s_y\right\} \sigma_0 
\nonumber \\ && \hspace{10mm} 
+ f_{xyz}({\bm k})s_z \sigma_x - d_{xz}({\bm k})s_z \sigma_y \bigr] i s_y, 
\end{eqnarray}
where $s_\alpha$ and $\sigma_\alpha$ are Pauli matrix in the spin and sublattice space, respectively. 
The orbital functions, $p_i({\bm k})$, $d_i({\bm k})$, and $f_i({\bm k})$, are obtained 
by assuming short-range Cooper pairs on neighboring ${\bm r}_i$ and ${\bm e}_i$ bonds~\cite{Supplemental1}. 
This choice is consistent with empirical rules obtained by microscopic calculations 
for many unconventional superconductors~\cite{Yanase_review}.
We choose $|\delta| \ll 1$ to study a dominantly $f$-wave state. 
Two-component order parameters are parametrized as 
$
(\eta_1, \eta_2) = \Delta (1,i \eta)/\sqrt{1+\eta^2} 
$
by a real parameter $\eta$. 
Ratio of $\eta_1$ and $\eta_2$ is pure-imaginary since the condensation energy is maximally gained in the 
chiral superconducting state. Thus, the B-phase is a chiral state 
where the range of $\eta$ is $0< \eta < \infty$~\cite{Supplemental3}. 
It is believed that a weak breakdown of hexagonal symmetry stabilizes 
the A- and C-phases~\cite{Sauls,Joynt,Aeppli,Hayden,Supplemental3}. 
We assume that the A-phase is the $\Gamma_2$-state ($\eta=\infty$), 
while the C-phase is the $\Gamma_1$-state ($\eta=0$).

The BdG Hamiltonian is represented in the Nambu space 
$
{\cal H}_{\rm BdG} = \frac{1}{2} \sum_{{\bm k}} \hat{c}_{\bm k}^\dagger \hat{H}_{\rm BdG}({\bm k}) \hat{c}_{\bm k},
$
with $\hat{c}_{\bm k} = \left(c_{{\bm k}1\uparrow},c_{{\bm k}2\uparrow},c_{{\bm k}1\downarrow},c_{{\bm k}2\downarrow} \right)^{\rm T}$. 
In order to study topological properties, we perform a unitary transformation
$
\tilde{H}_{\rm BdG}({\bm k}) =
U({\bm k}) \hat{H}_{\rm BdG}({\bm k}) U({\bm k})^{\dag}.  
$
By choosing 
$
U({\bm k}) = 
\left(
\begin{array}{cc}
1 & 0 \\
0 & e^{i{\bm k} \cdot {\bm \tau}} \\
\end{array}
\right)_{\sigma}
\otimes
s_0
\otimes
\tau_0 
$
and ${\bm \tau}=(0,-\frac{1}{\sqrt{3}},\frac{1}{2})$, $\tilde{H}_{\rm BdG}({\bm k})$ is periodic with respect to 
the translation ${\bm k} \rightarrow {\bm k} + {\bm K}$ with ${\bm K}$ being a reciprocal lattice vector.

Weyl nodes are specified by a topological Weyl charge defined by a monopole of Berry flux, 
$
q_i = \frac{1}{2\pi} \oint_{S} {\rm d}{\bm k} \vec{F}({\bm k}), 
$
where the Berry flux 
\begin{eqnarray}
F_{i}({\bm k}) = -i \varepsilon^{ijk}\sum_{E_n(\k)<0} \partial_{k_j} \langle u_n(\k)| \partial_{k_k} u_n(\k)\rangle, 
  \label{Berry_flux}
\end{eqnarray}
is integrated on a closed surface surrounding an isolated point node. 
We identify Weyl nodes by calculating $k_z$-dependent Chern number, 
\begin{eqnarray}
&&
\nu(k_z) = \frac{1}{2\pi} \int {\rm d}{\bm k}_\parallel F_{z}({\bm k}), 
\label{Chern}
\end{eqnarray}
on a two-dimensional $k_x$-$k_y$-plane~\cite{Thouless,Kohmoto,Fukui}. 
A wave function and energy of Bogoliubov quasiparticles are denoted by $|u_n(\k)\rangle$ and $E_n(\k)$, respectively. 
A jump in the Chern number is equivalent to the sum of Weyl charges at $k_z$. 
That is,
$
\nu(k_z + 0) -\nu(k_z - 0) = \sum_{i} q_i.
$
Thus, counting symmetry-related point nodes and comparing it with a jump in $\nu(k_z)$, 
we are able to identify Weyl charges.

\begin{figure}[htbp]
\begin{center}
\includegraphics[width=75mm]{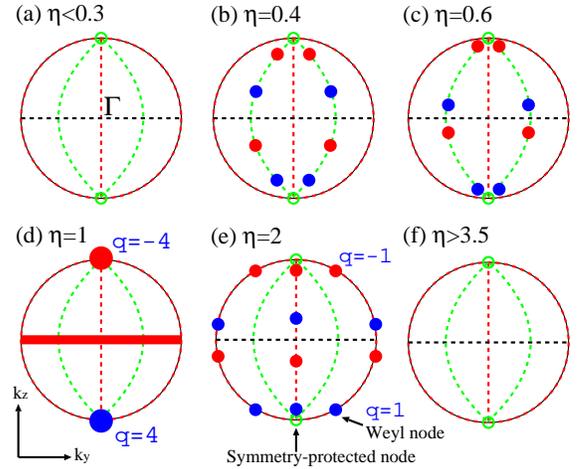}
\caption{(Color online)
Illustration of pair creation and annihilation of Weyl nodes on a $\Gamma$-FS. 
Projection from the $k_x$-axis is shown. 
(b), (c) and (e), blue and red closed circles show single Weyl nodes with $q_i = 1$ and $-1$, respectively. 
Large circles in (d) are spin-degenerate double Weyl nodes with $q_i = \pm 4$. 
Thick solid line in (d) shows a quadratic line node at $k_z = 0$. 
Green open circles are trivial point nodes protected by symmetry. 
Dashed lines illustrate nodal lines in the purely $f$-wave states, although they disappear 
in a generic $E_{\rm 2u}$-state. 
} 
\label{eGnode}
\end{center}
\end{figure}

\begin{figure}[htbp]
\begin{center}
\includegraphics[width=85mm]{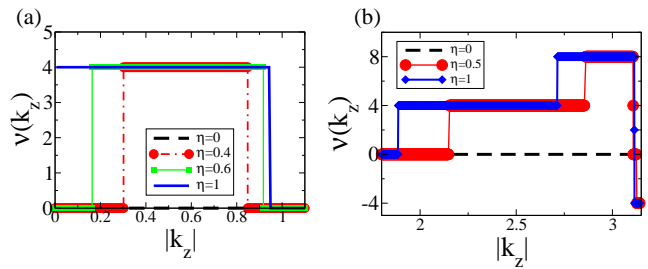}
\caption{(Color online)
Chern number of the two-dimensional BdG Hamiltonian parametrized by $k_z$ for (a) a $\Gamma$-FS reproduced by the parameter set 
$(t,t_z,t',\alpha,\mu,\Delta,\delta)=(1,4,1,0,16.0.4,0.01)$ and for 
(b) $A$-FSs by $(1,-4,1,0,12,0.1,0.04)$. 
} 
\label{kz-Chern}
\end{center}
\end{figure}

First, we elucidate Weyl nodes on the $\Gamma$-FS. This is a simple case, 
because only one of the bands crosses the Fermi level although the two-band model is adopted.
The nodal structures are qualitatively the same as those in the single-band model which are analytically expressed 
in appendix~\cite{Supplemental2}. 
Then, the $d$-wave order parameter does not play any important role.
On the other hand, the $p$-$f$ mixing in the order parameter eliminates line nodes in the chiral $f$-wave state~\cite{Sauls} 
in accordance with Blount's theorem~\cite{Blount,Kobayashi-Sato}, except for $\eta=1$, 
Diagonalizing $\tilde{H}_{\rm BdG}({\bm k})$,  
we obtain the superconducting gap illustrated in Fig.~\ref{eGnode}. 
Instead of line nodes, Weyl nodes (closed circles in Fig.~2) appear in the B-phase, in addition to the symmetry-protected 
point nodes at the poles on the FS. 
The former is identified as Weyl nodes by Fig.~\ref{kz-Chern}(a). 
The Chern number jumps by $\pm 4$, and we find four point nodes at a certain $k_z$. 
Thus, the point nodes are identified as single Weyl nodes with a unit charge $q_i = \pm 1$. 
In total, eight pairs of single Weyl nodes are produced. 
Since the spin degeneracy is lifted owing to the non-unitary order parameter, the spinless single Weyl nodes are realized 
even at zero magnetic field.

Here we show pair creation, annihilation and coalescence of Weyl nodes. 
Because the particle-hole symmetry is implemented in the BdG Hamiltonian for superconductors, the time-reversal symmetry 
ensures the chiral symmetry prohibiting Weyl nodes~\cite{Comment4}. 
Therefore, the A- and C-phases do not host Weyl nodes as illustrated in Figs.~\ref{eGnode}(a) and (f). 
The triviality is robust against a weak time-reversal-symmetry-breaking 
because the gap-closing is required for the topological transition. 
Therefore, pair creation of Weyl nodes does not occur at the thermodynamical A-B and B-C phase boundaries,  
but occurs in the B-phase. 
For the parameters in Fig.~\ref{kz-Chern}(a), we see the pair creation at $\eta= 0.3$ and $3.5$. 
Accordingly, the Weyl superconducting phase is illustrated in Fig.~1(a).

When the parameter $\eta$ is increased from zero by decreasing the magnetic field or increasing the temperature, 
the Weyl nodes emerge and move along nodal lines of the $f_{(x^2-y^2)z}$-wave component [Figs.~\ref{eGnode}(b) and (c)]. 
At $\eta=1$, four pairs of Weyl nodes cause pair annihilation at $k_z =0$, and remaining eight Weyl nodes 
coalesce into a pair of spin-degenerate double Weyl nodes ($q_i = \pm 4$) at the poles [Fig.~\ref{eGnode}(d)]. 
When $\eta$ increases from unity, eight pairs of Weyl nodes again appear on nodal lines of 
the $f_{xyz}$-wave component [Fig.~\ref{eGnode}(e)]. 
These pair creation, pair annihilation, and coalescence of Weyl nodes occur in the generic $E_{\rm 2u}$-state 
as a result of the $p$-$f$ mixing~\cite{Supplemental2}, although the chiral $f$-wave state hosts only Weyl nodes at the poles 
which are shown in Fig.~2(d)~\cite{Goswami}.

The pair annihilation and coalescence of Weyl nodes give rise to unusual gap structure at $\eta=1$. 
Since the B-phase is weakly non-unitary due to small $\delta$,
two non-equivalent superconducting gaps $\Delta_\pm({\bm k})$ are obtained~\cite{Sigrist-Ueda}. 
We see an intriguing nodal structure in the small gap $\Delta_-({\bm k})$. 
The pair annihilation leads to {\it quadratic line node} at $k_z=0$, 
that is, $\Delta_-({\bm k}) \propto |k_z|^2$, which is distinct from a usual linear line node 
with $\Delta({\bm k}) \propto |k_z|$ and gives rise to a low-energy DOS, $\rho(\omega) \propto \sqrt{\omega}$. 
On the other hand, the coalescence of Weyl nodes results in {\it cubic point nodes}, 
$\Delta_-({\bm k}) \propto |{\bm k}_\parallel|^3$, at the poles. 


Next, we investigate the $A$-FSs. This is an intriguing case, because the nonsymmorphic crystal symmetry 
causes gap nodes. 
The inter-sublattice hybridyzation  $a({\bm k})$ vanishes at $k_z = \pi$, and resulting sublattice degeneracy 
leads to paired FSs, as shown in Fig.~\ref{arc_node}(a). 
Although the degeneracy is partly lifted by the SOC, the symmetry protects the degeneracy along 
${\bm k}_\parallel \parallel $[010] and symmetric lines ($A$-$L$ lines)~\cite{Comment3}. 
Thus, any single band model breaks down, and our two-band model is a minimal model.

\begin{figure}[htbp]
\begin{center}
\includegraphics[width=80mm]{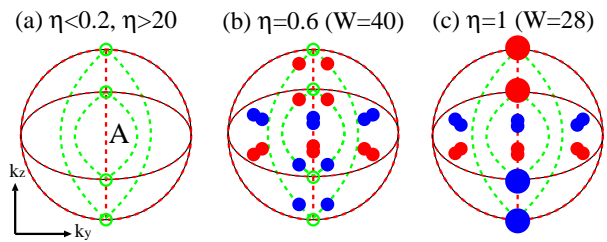}
\caption{(Color online)
Illustration of Weyl nodes on the $A$-FSs drawn by numerically diagonalizing 
the BdG Hamiltonian for the parameter set in Fig.~\ref{kz-Chern}(b). 
Thin solid lines show paired FSs. 
Other marks are the same as Fig.~\ref{eGnode}. 
We show the number of Weyl nodes $W$. 
} 
\label{Znode}
\end{center}
\end{figure}

Figure~\ref{kz-Chern}(b) shows an increase in the Chern number $\nu(k_z) = 0 \rightarrow 4 \rightarrow 8$ with $|k_z|$, 
indicating eight pairs of single Weyl nodes. 
These Weyl nodes arise from the $p$-$f$-mixing as we clarified for the $\Gamma$-FS. 
The Chern number $\nu(k_z) = 8$ is obtained just by the multiplication $2$ due to the two-bands. 
When we furthermore increase $|k_z|$ to $\pi$, interestingly the Chern number changes to $-4$ given 
by the $d$-wave component of the order parameter. 
Since the sublattice-singlet $d$-wave component induces inter-band pairing, it is negligible in most region of 
the Brillouin zone. However, the inter-band pairing may play an important role around $k_z =\pi$ 
where the two bands are nearly degenerate. 
The jump of Chern number, $\nu(k_z) = 8 \rightarrow -4$, results in twelve pairs of single Weyl nodes. 
Twenty pairs of Weyl nodes appear on the $A$-FSs in total [Fig.~4(b)]. 
In contrast to eight pairs due to the $p$-$f$-mixing, the twelve pairs of Weyl nodes arising from the $d$-$f$-mixing 
in the order parameter are robust in the six-fold rotation-symmetric state at $\eta=1$ [Fig.~4(c)], 
because twelve is a multiple of six.

\begin{figure}[htbp]
\begin{center}
\hspace*{-10mm}
\includegraphics[width=90mm]{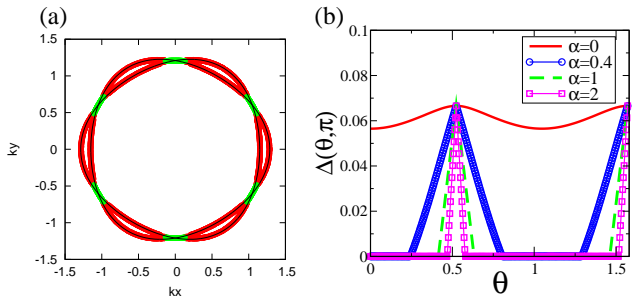}
\caption{(Color online)
(a) $A$-FSs at the Brillouin zone face, $k_z= \pm \pi$. Gapless and gapped regions at $\alpha=1$ 
are shown by red and green lines, respectively. 
(b) Angle dependence of the gap $\Delta(\theta,\pi)$ for various SOCs. 
The other parameters are the same as Fig.~3(b). 
} 
\label{arc_node}
\end{center}
\end{figure}

Now we show a gap node induced by the SOC. 
The nonsymmorphic space group allows a line node at the Brillouin zone face 
as pointed out by Norman~\cite{Norman} as a counterexample of Blount's theorem. 
Contrary to the Norman's argument, we obtain a {\it nodal arc} on the FSs. That is, a part of FS is gapped 
as shown in Fig.~\ref{arc_node}.  
Seemingly contradictory results are obtained because 
we plot the excitation gap near the FSs, $\Delta(\theta,\pi) = {\rm Min}_{k,n} |E_n(k \cos\theta, k \sin\theta,\pi)|$, 
although Norman's argument revealed the disappearance of intra-band Cooper pairs. 
The excitation spectrum is actually gapped around ${\bm k}_\parallel \parallel$[010] 
owing to the inter-band Cooper pairing. 

\begin{figure}[htbp]
\hspace*{-10mm}
\includegraphics[width=85mm]{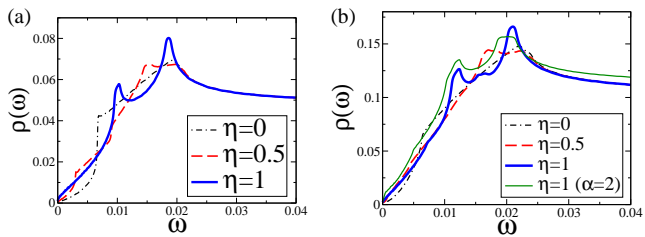}
\caption{(Color online)
DOS of (a) a $\Gamma$-FS and (b) $A$-FSs. 
Parameters are the same as Fig.~3. 
We also show a result for $\alpha=2$ in (b) by the thin green solid line in order to clarify the contribution 
of the SOC-induced arc node. 
} 
\label{dos}
\end{figure}

Finally, we show the DOS in Fig.~6. 
Even though the line node is absent in accordance with Blount's theorem, 
the DOS shows a linear energy dependence in the B-phase in agreement with experimental observations~\cite{Sauls,Joynt}. 
Although the SOC-induced arc node on the $A$-FSs~\cite{Norman} increases the DOS around $\omega=0$, 
its contribution is not dominant, as shown in Fig.~6(b). Thus, it is indicated that 
at most 44 point nodes on paired A-FSs and 54 point nodes on three $\Gamma$-FSs give rise to line node-like behaviors. 
The thermal conductivity measurement has been a powerful tool in identifying the superconducting symmetry 
of UPt$_3$~\cite{Joynt,Machida-Izawa}. 
The saturating ratio $\kappa_c/\kappa_b$ supporting the chiral $f$-wave state~\cite{Joynt,Lussier,Norman-Hirschfeld} 
also supports the generic $E_{\rm 2u}$-state which has point nodes on general points of Brillouin zone. 
A residual thermal conductivity much smaller than the universal value of line nodal superconductors~\cite{Suderow} 
may be consistent with point nodal 
$E_{\rm 2u}$-state. Nearly isotropic {\it ab}-plane field-angle dependence~\cite{Machida-Izawa,Kittaka_2013} 
may be explained by the cancellation of anisotropy from nearly $100$ point nodes.

A square-root dependence $\rho(\omega) \propto \sqrt{\omega}$ is obtained in the low-energy region at $\eta=1$ 
[Fig.~6(a)] as a result of the quadratic line node. 
However, unusual $\sqrt{\omega}$-dependence may be obscured by the vortex scattering 
since $\eta=1$ is not realized at zero magnetic field as illustrated in Fig.~1(a)~\cite{Supplemental3}.

{\it Conclusion ---} 
To conclude, the $E_{\rm 2u}$-pairing-state in UPt$_3$ is a Weyl superconducting state of new type 
which shows the pair creation, pair annihilation, and coalescence of Weyl nodes. 
The $p$+$d$+$f$-mixing in the order parameter partly due to the nonsymmorphic crystal symmetry causes 
these unusual behaviors which have not been observed in a chiral $f$-wave state~\cite{Goswami}. 
The topologically distinct properties give rise to Majorana arcs in surface states analogous 
to recently observed Fermi arc~\cite{Xu,Lv,Yang}, resulting 
in a zero-field thermal Hall conductivity~\cite{Fischer_Weyl,Goswami} and quasiparticle interference~\cite{Kourtis}. 
Tunable positions of Weyl nodes by temperature and magnetic field may enable experimental observations 
and also may induce the chiral anomaly through a topological defect in the combined real and 
momentum space~\cite{Volovik_chiral_anomaly}.

Our results are compatible with Blount's theorem~\cite{Blount,Kobayashi-Sato}, but at most 98 point nodes 
and the SOC-induced arc node lead to line node behaviors in the DOS, 
consistent with experimental observations in UPt$_3$~\cite{Joynt}. 
A generic $E_{\rm 2u}$-pairing-state studied here may also be consistent with experiments incompatible 
with the chiral $f$-wave state~\cite{Machida-Izawa,Tou_UPt3,Suderow,Kittaka_2013}, although further theoretical developments taking account of  
multigap structure are desired.  

Nonsymmorphic symmetry may be weakly broken either by a crystal distortion~\cite{Walko} 
or by an antiferromagnetic order in UPt$_3$~\cite{Aeppli,Hayden}. 
Even in this case, the topologically protected Weyl nodes are robust, 
although the symmetry-protected SOC-induced arc node is gapped.

\begin{acknowledgments}
The authors are grateful to K. Izawa, H. Harima, S. Kobayashi, T. Nomoto, M. Sato, and K. Shiozaki 
for fruitful discussions. 
This work was supported by the ``Topological Quantum Phenomena'' (No. JP25103711) and ``J-Physics'' 
(No. JP15H05884) Grant-in Aid for Scientific Research on Innovative Areas from MEXT of Japan, 
and by JSPS KAKENHI Grant Nos. JP24740230, JP15K05164, JP15H05745, and JP16H00991. 
\end{acknowledgments}

\bibliography{reference}

\begin{thebibliography}{33}
\bibitem{Comment1}
A line node protected by a nonsymmorphic space group symmetry has been proposed~\cite{Norman} as 
a counterexample of Blount's theorem. 
Indeed, our model shows an arc node protected by the screw and glide symmetry of UPt$_3$. 
On the other hand, the line node in the chiral $f$-wave state is not protected 
by the nonsymmorphic symmetries.
\bibitem{Supplemental3}
The range of the parameter $\eta$ is listed in Appendix C (Table.~II). 
The temperature and magnetic field dependence of the parameter $\eta$ is explained 
in details on the basis of the Ginzburg-Landau theory.
\bibitem{Supplemental2}
Analytic results of nodal structure in the single band model are shown in Appendix B.
\bibitem{Comment3}
The degeneracy of two bands at $k_z = \pi$ and ${\bm k_\parallel} \parallel $[010] ($A$-$L$ lines) 
is protected by the crystal symmetry. This is proved by using the relations, 
$\{IG_{xz},IT\}=0$, $\{IG_{xz},M_{yz}\}=0$, and $\left(IG_{xz}\right)^2=-1$, at $k_z =\pm \pi$.  
Inversion symmetry, time-reversal symmetry, and mirror symmetry are represented 
by $I=\sigma_x$, $T=is_y K$, and $M_{yz}= is_x$, respectively. 
The glide symmetry is represented by $G_{xz}=s_y \sigma_y$ at $k_z=\pm \pi$ 
while $G_{xz}=is_y \sigma_x$ at $k_z=0$.
We will show details elsewhere.
\end{thebibliography}

\clearpage
\onecolumngrid

\appendix

\section{Order Parameter in $E_{\rm 2u}$ Representation of $D_{\rm 6h}$ Point Group with Nonsymmorphic Operation}

 In the presence of spin-orbit coupling, superconducting states are classified by crystal 
point group symmetry. The order parameter should belong to an irreducible representation of 
the point group~\cite{Sigrist-Ueda}. 
In this paper we adopt the order parameter of superconductivity in the $E_{\rm 2u}$ representation 
of $D_{\rm 6h}$ point group symmetry. List of characters of the two-dimensional $E_{\rm 2u}$ representation 
is shown in Table~\ref{SM_tab1}. 
Basis functions are composed of the $p$-wave component, $\left(\hat{\Gamma}_1, \hat{\Gamma}_2\right) = 
\left( p_{x} \hat{x} -p_{y} \hat{y}, p_{y} \hat{x} + p_{x} \hat{y}\right)$, and the $f$-wave component, 
$\left(f_{(x^2-y^2)z} \hat{z}, f_{xyz} \hat{z} \right)$, 
in the absence of the orbital and sublattice degrees of freedom~\cite{Joynt}. 

On the other hand, the crystal structure of UPt$_3$ is composed of two sublattices at $z=0$ and $z=1/2$. 
The local symmetry of Uranium ions is $D_{\rm 3h}$ which lacks inversion symmetry, while 
the stacking of Uranium ions along the $c$ axis recovers the inversion symmetry and 
ensures the point group $D_{\rm 6h}$. 
In this crystal structure, a half of symmetry operations of the $D_{\rm 6h}$ point group, 
which are not included in the $D_{\rm 3h}$ point group, is accompanied by a half translation 
along the $c$ axis. For instance, glide symmetry $\{\sigma_{\rm v}|\frac{\hat{c}}{2}\}$ 
and screw symmetry $\{C_6|\frac{\hat{c}}{2}\}$ are preserved instead of $\sigma_{\rm v}$ 
mirror symmetry and $C_6$ rotation symmetry. Thus, the space group is nonsymmorphic. 
Then, we can show that the sublattice-singlet spin-triplet $d$-wave component, 
$\left(\hat{\Gamma}_1, \hat{\Gamma}_2\right) = 
\left( d_{yz} \hat{z}, d_{xz} \hat{z} \right)\sigma_y$,
is also a basis function of the $E_{\rm 2u}$ representation. 
Indeed, the symmetry operations on the $d$-wave component are specified by the characters 
of the $E_{\rm 2u}$ representation as we show in Table~\ref{SM_tab1}, although the conventional (sublattice-triplet spin-singlet) 
$d$-wave state belongs to the $E_{\rm 1g}$ representation.

\newcolumntype{C}{>{\centering}p{4cm}}
\begin{table}[htbp]
  \begin{tabular}{c|cccccc|cccccc||C}
    \hline\hline
 & $E$ & $2C_3$ & $3C_2'$ & $2S_3$ & $\sigma_{\rm h}$ & $3 \sigma_{\rm d}$ 
& $2\{C_6|\frac{\hat{c}}{2}\}$ & $\{C_2|\frac{\hat{c}}{2}\}$ & $3C_2''$ & $I$ & $S_6$ & 
$3\{\sigma_{\rm v}|\frac{\hat{c}}{2}\}$ & Basis function 
\tabularnewline \hline
$E_{\rm 2u}$ & 2 & -1 & 0 & 1 & -2 & 0 & -1 & 2 & 0 & -2 & 1 & 0 & 
$(k_x \hat{x} -k_y \hat{y}, k_y \hat{x} +k_x \hat{y})$, \\
$(k_x^2-k_y^2, 2k_x k_y) k_z \hat{z}$, \\
$(k_y k_z, k_z k_x)\hat{z}\sigma_y$ 
\tabularnewline \hline
$E_{\rm 1g}$ & 2 & -1 & 0 & 1 & -2 & 0 & 1 & -2 & 0 & 2 & -1 & 0 & $(k_z k_x, k_y k_z)$
\tabularnewline \hline\hline
  \end{tabular}
  \caption{List of characters for the $E_{\rm 2u}$ and $E_{\rm 1g}$ representations of the $D_{\rm 6h}$ point group. 
The last column shows basis functions.}
  \label{SM_tab1}
\end{table}

One may suppose that the sublattice-singlet spin-triplet $d$-wave component requires an exotic Cooper pairing. 
However, it naturally appears when we adopt short-range Cooper pairs. 
We take into account the nearest-neighboring Cooper pairs formed on the ${\bm r}_i$-bonds, 
that is shown in Fig.~1(b). 
The pairing field by each bond is obtained as, $\Delta_i  \sin\frac{k_z}{2} e^{i {\bm k}_\parallel \cdot {\bm r}_i}$, 
with $i=1,2,3$. The order parameter compatible with $E_{\rm 2u}$ irreducible representation is given by 
\begin{align}
\hat{\Delta}_{d+f}({\bm k}) &= 
\left(
\begin{array}{cc}
0 & \sum_i \Delta_i  \sin\frac{k_z}{2} e^{i {\bm k}_\parallel \cdot {\bm r}_i}
\\ 
\sum_i \Delta_i  \sin\frac{k_z}{2} e^{-i {\bm k}_\parallel \cdot {\bm r}_i} & 0
\end{array}
\right)_\sigma \otimes  i s_z s_y. 
\label{SM_d+f-wave}
\end{align}
The real part proportional to $\sigma_x$ may have $f$-wave symmetry. 
Taking appropriate linear combination, we obtain the orbital functions of $f_{xyz}$-wave and $f_{(x^2-y^2)z}$-wave 
components, 
\begin{eqnarray}
f_{xyz}({\bm k}) = -\sqrt{3} \sin\frac{k_z}{2} {\rm Re} \sum_{i} r_i^{x} e^{i {\bm k}_\parallel \cdot {\bm r}_i},  
\label{SM_f1}
\\
f_{(x^2-y^2)z}({\bm k}) = -\sqrt{3} \sin\frac{k_z}{2} {\rm Re} \sum_{i} r_i^{y} e^{i {\bm k}_\parallel \cdot {\bm r}_i}.  
\label{SM_f2}
\end{eqnarray}
The imaginary part of Eq.~(\ref{SM_d+f-wave}) is represented by the Pauli matrix $\sigma_y$ and 
the symmetry is $d$-wave. 
The same linear combination as Eqs.~(\ref{SM_f1}) and (\ref{SM_f2}) gives the orbital functions of $d_{xz}$-wave 
and $d_{yz}$-wave components, 
\begin{eqnarray}
d_{xz}({\bm k}) = -\sqrt{3} \sin\frac{k_z}{2} {\rm Im} \sum_{i} r_i^{x} e^{i {\bm k}_\parallel \cdot {\bm r}_i}, 
\label{SM_d1}
\\
d_{yz}({\bm k}) = -\sqrt{3} \sin\frac{k_z}{2} {\rm Im} \sum_{i} r_i^{y} e^{i {\bm k}_\parallel \cdot {\bm r}_i}. 
\label{SM_d2}
\end{eqnarray}
By using these orbital functions, the order parameter of Cooper pairs on the ${\bm r}_i$-bonds is written as 
\begin{align}
\hat{\Delta}_{d+f}({\bm k}) &=
\eta_1 \bigl[  f_{(x^2-y^2)z}({\bm k}) \sigma_x - d_{yz}({\bm k}) \sigma_y \bigr] s_z i s_y
+ \eta_2 \bigl[  f_{xyz}({\bm k}) \sigma_x - d_{xz}({\bm k}) \sigma_y \bigr] s_z i s_y. 
\label{SM_d+f-wave2}
\end{align}
Thus, the $d$-wave component naturally accompanies with the $f$-wave component as a result of 
the crystal structure of UPt$_3$. In other words, the Cooper pairs on the ${\bm r}_i$-bonds give rise to 
superconducting order parameters having composite $d$+$f$-wave momentum dependence. 
Equation~(\ref{SM_d+f-wave2}) is a part of Eqs.~(2) and (3) in the main text.

Because the $p$-wave component also belongs to the $E_{\rm 2u}$ irreducible representation, 
it naturally coexists with the above $d+f$-wave component owing to the spin-orbit coupling. 
In this paper we adopt $p$-wave Cooper pairs formed on the intra-layer nearest-neighboring  
${\bm e}_i$-bonds. Then, the order parameter compatible with $E_{\rm 2u}$ irreducible representation is 
\begin{align}
\hat{\Delta}_p({\bm k}) &= 
\eta_1 \bigl[p_x({\bm k})s_x - p_y({\bm k})s_y\bigr] \sigma_0 i s_y 
+\eta_2 \bigl[p_y({\bm k})s_x + p_x({\bm k})s_y\bigr] \sigma_0 i s_y, 
\label{SM_p-wave}
\end{align}
and the orbital functions are given by 
\begin{eqnarray}
p_{x}({\bm k}) = \sum_{i} e_i^{x}  \sin {\bm k}_\parallel\cdot{\bm e}_i, 
\label{SM_p1}
\\
p_{y}({\bm k}) = \sum_{i} e_i^{y}  \sin {\bm k}_\parallel\cdot{\bm e}_i. 
\label{SM_p2}
\end{eqnarray}
Taking into account the mixing of the $d+f$-wave Cooper pairs on the ${\bm r}_i$-bonds and 
the $p$-wave Cooper pairs on the ${\bm e}_i$-bonds, we adopt a linear combination 
$\hat{\Delta}({\bm k}) = \delta \hat{\Delta}_p({\bm k}) + \hat{\Delta}_{d+f}({\bm k})$. 
Then, the basis functions of generic $E_{\rm 2u}$ state are given by 
Eqs.~(2) and (3) in the main text. 
To be compatible with experiments, we assume that the weight of $p$-wave component is small, $|\delta| \ll 1$.

Recently, the mixing of $p$-wave, $d$-wave, and $f$-wave order parameters was demonstrated 
by a microscopic theory based on first principle band structure calculation~\cite{Nomoto}. 
Although the weight of $p$-, $d$-, and $f$-wave components obtained in Ref.~\cite{Nomoto} 
is unlikely to be compatible 
with experiments~\cite{Joynt}, it is indicated that mixing of order parameters naturally occurs 
owing to the many body effect. Generally speaking, symmetry-allowed interactions effectively appear 
in solid state electron systems through the many body effect. Therefore, symmetry-allowed mixing 
in order parameter naturally occurs.

It should be noticed that the inter-sublattice $d$+$f$-wave Cooper pairs and the intra-sublattice $p$-wave ones 
are taken into account.  
Intra-sublattice $f$-wave Cooper pairs as well as inter-sublattice $p$-wave ones may also be induced 
by a longer-range pairing. However, we have confirmed that those components do not alter our results 
as far as the $d$+$f$-wave component is a dominant order parameter.

Note that the short-range Cooper pairs have been identified in many unconventional superconductors 
by microscopic theories for the strongly correlated electron systems~\cite{Yanase_review}. 
Thus, the order parameter derived above is a reasonable assumption. 
Although it is desirable to calculate the order parameter with the use of microscopic theories, 
satisfactory theory has not been developed for Uranium-based heavy fermion superconductors up to now.

\section{Nodal structure in a single band model}

Any single band model does not reproduce the nodal structure on the paired $A$-FSs illustrated in Figs.~4 and 5, 
because the nonsymmorphic crystal symmetry and associated multiband structure are not taken into account. 
On the other hand, the Weyl node, quadratic line node, and cubic point node on the $\Gamma$-FS (Fig.~2) 
are obtained even in the single band model. 
These unusual nodal structures are indeed induced by the $p$-$f$-mixing in the order parameter, 
and then the multiband structure and spin-orbit coupling do not play any role. 
Analytic results of nodal structures are obtained in the single band model, which we show below 
for a transparent understanding of Fig.~2.

We consider the basis function composed of the $p$-wave and $f$-wave components,  
\begin{eqnarray}
&&
\hat{\Gamma}_1 = \bigl[\delta \left\{p_x({\bm k})s_x - p_y({\bm k})s_y\right\} 
+ f_{(x^2-y^2)z}({\bm k})s_z \bigr] i s_y,
\\
&&
\hat{\Gamma}_2 = \bigl[\delta \left\{p_y({\bm k})s_x + p_x({\bm k})s_y\right\} 
+ f_{xyz}({\bm k})s_z \bigr] i s_y. 
\end{eqnarray}
The order parameter in the chiral superconducting state,  
$
\hat{\Delta}({\bm k}) = \Delta_0 \left( \hat{\Gamma}_1 + i \eta \hat{\Gamma}_2 \right)  
$, 
is described by the $d$-vector 
\begin{eqnarray}
{\bm d}({\bm k}) = \Delta_0 \left[ 
\delta \left\{p_x({\bm k})\hat{x} - p_y({\bm k})\hat{y} \right\} 
+ i \eta \delta \left\{p_y({\bm k})\hat{x} + p_x({\bm k})\hat{y} \right\} 
+ f_{(x^2-y^2)z}({\bm k}) \hat{z} + i \eta f_{xyz}({\bm k}) \hat{z}
\right]. 
\nonumber \\
\end{eqnarray}
Because the order parameter is non-unitary~\cite{Sigrist-Ueda}, the superconducting gap $\Delta_\pm({\bm k})$ is obtained by 
\begin{eqnarray}
\Delta_\pm({\bm k})^2 &=&  |{\bm d}({\bm k})|^2 \pm |i {\bm d}({\bm k}) \times {\bm d}({\bm k})^*|^2
\\
&=& \Delta_0^2 \left[ \left(1 + \eta^2\right) |p({\bm k})|^2 + |f_{(x^2-y^2)z}({\bm k})|^2 + \eta^2 |f_{xyz}({\bm k})|^2 
\pm 2\sqrt{\eta^2 |p({\bm k})|^2  \left(|p({\bm k})|^2+|f({\bm k})|^2 \right)} \right], 
\nonumber \\
\label{gap_single}
\end{eqnarray}
with $|p({\bm k})|^2 = \delta^2 \left(|p_x({\bm k})|^2 + |p_y({\bm k})|^2 \right) $ 
and $|f({\bm k})|^2 = \left(|f_{(x^2-y^2)z}({\bm k})|^2 + |f_{xyz}({\bm k})|^2 \right) $. 
For $\eta \leq 1$, the condition for a nodal gap, $\Delta_-({\bm k})=0$, is satisfied when 
\begin{eqnarray}
f_{(x^2-y^2)z}({\bm k})=0 
\hspace{3mm}
{\rm and}
\hspace{3mm}
\left(1-\eta^2 \right)|p({\bm k})|^2 = \eta^2 |f_{xyz}({\bm k})|^2. 
\label{node_single1}
\end{eqnarray}
On the other hand, for $\eta \geq 1$ the small gap $\Delta_-({\bm k})$ may have nodes at ${\bm k}$ with 
\begin{eqnarray}
f_{xyz}({\bm k})=0 
\hspace{3mm}
{\rm and}
\hspace{3mm}
\left(\eta^2 -1 \right)|p({\bm k})|^2 = |f_{(x^2-y^2)z}({\bm k})|^2. 
\label{node_single2}
\end{eqnarray}
Hence, the nodal structure is clarified by solving Eqs.~(\ref{node_single1}) and (\ref{node_single2}).

Let us assume an isotropic Fermi surface at 
${\bm k} = k_{\rm F}(\sin\theta \cos\phi, \sin\theta \sin\phi, \cos\theta)$ 
for simplicity. Then, we naturally adopt the long wave length approximation, 
\begin{eqnarray}
&& \hspace{-5mm}
\left[p_x({\bm k}),p_y({\bm k})\right] = (k_x,k_y)/k_{\rm F} = k \sin\theta (\cos\phi, \sin\phi)/k_{\rm F}, 
\\
&& \hspace{-5mm}
\left[f_{(x^2-y^2)z}({\bm k}),f_{xyz}({\bm k})\right] = (k_x^2 - k_y^2,2 k_x k_y)k_z/k_{\rm F}^3 
= k^3 \sin^2\theta \cos\theta (\cos2\phi, \sin2\phi)/k_{\rm F}^3. 
\end{eqnarray}
Solving Eq.~(\ref{node_single1}), we find point nodes at $\phi=n\pi/2+\pi/4$ with $n$ being an integer and 
\begin{eqnarray}
&& \hspace{-8mm}
\theta = 
\pi/4 \pm \left[\pi/4 - \frac{1}{2} \arcsin\left(2\delta\sqrt{\eta^{-2}-1} \right) \right], 
3 \pi/4 \pm \left[\pi/4 - \frac{1}{2} \arcsin\left(2\delta\sqrt{\eta^{-2}-1} \right) \right]. 
\end{eqnarray}
Equation~(\ref{node_single2}) is satisfied at the momentum, $\phi= n \pi/2$ and 
\begin{eqnarray}
&& \hspace{-8mm}
\theta = 
\pi/4 \pm \left[\pi/4 - \frac{1}{2} \arcsin\left(2\delta\sqrt{\eta^2-1} \right) \right], 
3 \pi/4 \pm \left[\pi/4 - \frac{1}{2} \arcsin\left(2\delta\sqrt{\eta^2-1} \right) \right]. 
\end{eqnarray}
From these results we find that the pair creation of Weyl nodes occurs 
at momentum ${\bm k} = k_{\rm F}\left(\pm \frac{1}{2}, \pm \frac{1}{2},\pm \frac{1}{\sqrt{2}}\right)$ 
when $\eta = \eta_{\rm c} \equiv \left(1+1/4\delta^2 \right)^{-1/2}$ 
while at ${\bm k} = k_{\rm F}\left(\pm \frac{1}{\sqrt{2}}, 0, \pm \frac{1}{\sqrt{2}}\right)$ and 
$k_{\rm F}\left(0, \pm \frac{1}{\sqrt{2}}, \pm \frac{1}{\sqrt{2}}\right)$ 
when $\eta = \eta_{\rm c}^{-1}$. 
Thus, the superconducting state hosts eight pairs of Weyl nodes for $\eta_{\rm c} < \eta < \eta_{\rm c}^{-1}$. 
These features of Weyl nodes have been demonstrated in the two-band model reproducing the $\Gamma$-FS 
(Fig.~2 in the main text). 
Trivial point nodes are present at $\theta=0$ and $\pi$ independent of the parameter $\eta$, 
since $p_x({\bm k})=p_y({\bm k})=f_{(x^2-y^2)z}({\bm k})=f_{xyz}({\bm k})=0$ at ${\bm k}_\parallel = 0$. 

At $\eta=1$, Eqs.~(\ref{node_single1}) and (\ref{node_single2}) are satisfied at $k_z =0$, 
leading to the quadratic line node at the equator of Fermi surface. 
Then, the positions of Weyl nodes discussed above are reduced to $\theta=0$, $\pi/2$, and $\pi$, 
indicating the pair annihilation at the equator of Fermi surface and the coalescence at the poles. 
Considering rotation symmetry, we understand the pair annihilation and coalescence of Weyl nodes. 
Six-fold rotation symmetry is preserved at and only at $\eta=1$. Then, the number of Weyl nodes 
at ${\bm k}_\parallel \ne 0$ (defined as $W'$) has to be a multiple of six.  
Since $W'=16$ at $\eta \ne 1$, it must change at $\eta=1$. 
It actually changes to $W'=0$ as a result of the pair annihilation and coalescence of Weyl nodes.

At $\eta=1$, the quadratic line node at $k_z=0$ and the cubic point nodes at ${\bm k}_\parallel=0$ are obtained 
from Eq.~(\ref{gap_single}). Because the $f$-wave component vanishes at these momentum, we can approximate 
the superconducting gap around the nodal region, 
\begin{eqnarray}
\Delta_{-}({\bm k})^2 
&=&  \Delta_0^2 
\left[2 |p({\bm k})|^2 + |f({\bm k})|^2 - 2 \sqrt{|p({\bm k})|^2  \left(|p({\bm k})|^2+|f({\bm k})|^2 \right)} \right]
\simeq \Delta_0^2 \frac{|f({\bm k})|^4}{4|p({\bm k})|^2},
\\
& \propto & k_z^4 \hspace{8mm} {\rm around} \hspace{2mm} k_z=0, 
\label{quadratic_line_node}
\\
& \propto & |{\bm k}_\parallel|^6 \hspace{4mm} {\rm around} \hspace{2mm} {\bm k}_\parallel=0.  
\label{cubic_point_node}
\end{eqnarray}
Equation~(\ref{quadratic_line_node}) shows the quadratic line node, $\Delta_{-}({\bm k}) \propto k_z^2 $, at the equator, 
while Eq.~(\ref{cubic_point_node}) shows the cubic point node, $\Delta_{-}({\bm k}) \propto |{\bm k}_\parallel|^3 $, 
at the poles. These higher-order nodal structures are also obtained in the two-band model 
(see the discussion in the main text). 
From Eq.~(\ref{gap_single}) we also obtain the linear point node in the high-energy branch, 
$\Delta_{+}({\bm k}) \propto |{\bm k}_\parallel|$, at ${\bm k}_\parallel=0$. 

In the next section S3, we show that the superconducting state with $\eta=1$ 
is not realized at zero magnetic field on the basis of the Ginzburg-Landau theory.
Although the magnetic field may realize $\eta=1$ through the symmetry breaking term, 
the unusual nodal structures may be obscured in the vortex state.

\section{Ginzburg-Landau Theory and Range of Parameter $\eta$}

The order parameter of superconductivity in the $E_{2u}$ representation is formally represented by 
\begin{align}
\hat{\Delta}({\bm k}) = \eta_1 \hat{\Gamma}_1 + \eta_2 \hat{\Gamma}_2, 
\end{align}
and the basis functions $\hat{\Gamma}_1$ and $\hat{\Gamma}_2$ are shown in Sec.~S1. 
The two component order parameters $\eta_1$ and $\eta_2$ are parametrized by 
\begin{align}
(\eta_1, \eta_2) = \Delta (1,i \eta)/\sqrt{1+\eta^2}. 
\end{align}
The ratio $\eta$ characterizes internal structure of Cooper pairs, which depends on temperature and magnetic field.  
Previous studies on UPt$_3$ revealed the range of $\eta$ listed in Table~\ref{SM_tab2}~\cite{Joynt,Sauls}. 
The $\eta$ in the A- and C-phases may be $(\eta_{\rm A}, \eta_{\rm C})=(0,\infty)$ or 
$(\eta_{\rm A}, \eta_{\rm C})=(\infty, 0)$. We assume the latter without loss of generality. 
Then, the $\eta$ changes from $0$ to $\infty$ in the B-phase, while it is zero (infinite) in the C-phase (A-phase). 
\begin{table}[htbp]
  \begin{tabular}{c|c}
\hline
A-phase & $\eta=\infty$ 
\\ \hline
B-phase & $0 \le \eta \le \infty $ 
\\ \hline
C-phase & $\eta=0$ 
\\ 
 \hline
  \end{tabular}
  \caption{Range of the parameter $\eta$ in the A-, B-, and C-phases of UPt$_3$~\cite{Joynt,Sauls}. 
}
  \label{SM_tab2}
\end{table}

\begin{figure}[htbp]
\begin{center}
\includegraphics[width=75mm]{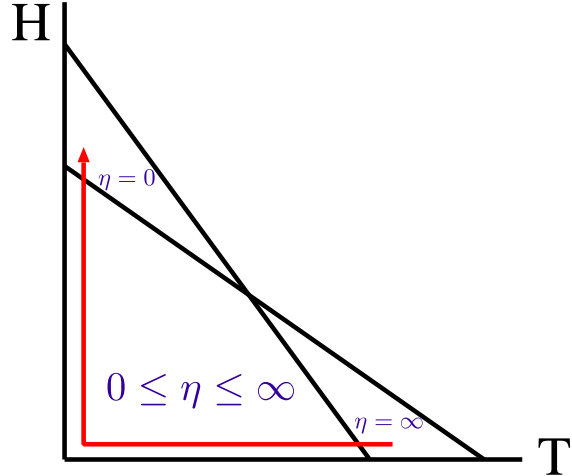}
\caption{(Color online) 
Illustration of the magnetic field-temperature phase diagram of UPt$_3$~\cite{Joynt}. 
The parameter $\eta$ in the A-, B-, and C-phases is shown. The A-phase with $\eta=\infty$ is stabilized by 
a symmetry breaking term $F_1$ which may be induced by a weak antiferromagnetic order. The C-phase with $\eta = 0$ 
is stabilized by another symmetry breaking term $F_1^{\rm grad}$ which renormalizes the coherence length. 
Because the A-B transition and B-C transition are second order phase transition and any first order transition does not occur in the B-phase, the parameter $\eta$ has to change from $\infty$ to $0$ when we move along the red arrow. 
} 
\label{schematicSM}
\end{center}
\end{figure}

Below we prove the list in Table~\ref{SM_tab2} on the basis of the Ginzburg-Landau theory. 
Since the A-B transition and B-C transition in the superconducting state are second order phase 
transition~\cite{Joynt,Sauls}, it is satisfactory to show the parameter $\eta$ in the A- and C-phases. 
The parameter $\eta$ must change continuously through the second order phase transitions, and therefore, 
it changes from $\infty$ to $0$ along the line in Fig.~\ref{schematicSM} drawing the $H$-$T$ phase diagram.
We also show that a special condition $\eta=1$ is not realized at zero magnetic field. 

The Ginzburg-Landau free energy density for chiral superconductors is given by $F_{0}+F_0^{\rm grad}$ with 
\begin{eqnarray}
  F_{0} &=& \alpha (|\eta_1 |^2+|\eta_2 |^2 ) 
  +\frac{\beta_1}{2} (|\eta_1 |^2+|\eta_2 |^2 )^2 
  +\frac{\beta_2}{2}(\eta_1 \eta_2^* - {\rm c.c.})^2 
  +\beta_3 |\eta_1 |^2 |\eta_2 |^2,   
  \label{eq:f0} 
\\
  F_0^{\rm grad} &=& 
   \kappa_1 ( |D_{\mathrm{x}}\eta_1 |^2 + |D_{\mathrm{y}}\eta_2 |^2 ) 
  +\kappa_2 ( |D_{\mathrm{x}}\eta_2 |^2 + |D_{\mathrm{y}}\eta_1 |^2 ) 
  +\kappa_5 ( |D_{\mathrm{z}}\eta_1 |^2 + |D_{\mathrm{z}}\eta_2 |^2 ) 
  \nonumber\\
  &&
  +\kappa_3 \bigl[(D_{\mathrm{x}}\eta_1 )(D_{\mathrm{y}}\eta_2 )^* + {\rm c.c.} \bigr] 
  +\kappa_4 \bigl[(D_{\mathrm{x}}\eta_2 )(D_{\mathrm{y}}\eta_1 )^* + {\rm c.c.} \bigr], 
  \label{eq:f0grad} 
\end{eqnarray}
for the two-component order parameters $(\eta_1, \eta_2)$ in the two-dimensional 
irreducible representation~\cite{Sigrist-Ueda,Joynt,Sauls},  
where $\alpha = \alpha_0 \left(\frac{T}{T_{\rm c0}}-1\right)$ and $D_j = -i \partial_j + (2\pi/\Phi_0)A_j$ 
are covariant derivatives. 
The hexagonal $D_{\rm 6h}$ point group symmetry imposes relations, $\beta_3=0$ and $\kappa_3 = \kappa_4 =(\kappa_1 - \kappa_2)/2$.

We can drop the gradient term at zero magnetic field and obtain the order parameters by minimizing 
the homogeneous term $F_0$. The superconducting state is determined by the sign of $\beta_2$. 
The chiral state, $(\eta_1, \eta_2) = \Delta_0 (\pm i, 1)$, is stable for $\beta_2 > 0$, while a non-chiral state, 
$(\eta_1, \eta_2) = \Delta_0 (\cos \theta,\sin \theta)$, is stable for $\beta_2 < 0$. 
The weak coupling BCS theory results in $\beta_2/\beta_1 =1/3$, and experimental results in UPt$_3$ are fitted by 
$\beta_2/\beta_1 = 0.16 \sim 0.33$~\cite{Sauls}. Thus, we assume $\beta_2 >0$. 

It is believed that the six-fold rotation symmetry is weakly broken in UPt$_3$~\cite{Sauls} 
possibly by the antiferromagnetic order~\cite{Aeppli,Hayden}. 
The symmetry breaking field yields a quadratic term,  
\begin{eqnarray}
F_1 &=& \alpha_0 \epsilon_0 \left(|\eta_1|^2 - |\eta_2|^2\right), 
\label{eq:f1}
\end{eqnarray}
and splits the transition temperature of the $\Gamma_1$-state and the $\Gamma_2$-state. 
Here $\epsilon_0 > 0$ is assumed without loss of generality. 
By minimizing $F_0 + F_1$, we obtain the transition temperature of the A-phase, 
$T_{\rm c}^{\rm A} = \left(1 + \epsilon_0 \right)T_{\rm c0}$, and the A-B transition temperature, 
$T_{\rm c}^{\rm B} = T_{\rm c}^{\rm A} - \frac{\beta_1}{\beta_2} \epsilon_0 T_{\rm c0}$.  
The order parameter in the A-phase is obtained as $\eta_1 =0$ and
\begin{eqnarray}
\eta_2= \sqrt{\frac{\alpha_0}{\beta_1 T_{\rm c0}} \left( T_{\rm c}^{\rm A}-T \right)}. 
\end{eqnarray}
Thus, the parameter $\eta=\infty$ is consistent with the list in Table~\ref{SM_tab2}. 

In the B-phase, $(\eta_1, \eta_2) = (\pm i |\eta_1|, |\eta_2|)$ and 
\begin{eqnarray}
\left(
\begin{array}{c}
|\eta_1|^2 \\
|\eta_2|^2
\end{array}
\right)
&=& 
- 
\left(
\begin{array}{cc}
\beta_1 & \beta_1 - 2 \beta_2 \\
\beta_1 - 2 \beta_2 & \beta_1 \\
\end{array}
\right)^{-1}
\left(
\begin{array}{c}
\alpha_- \\
\alpha_+
\end{array}
\right),
\end{eqnarray}
where $\alpha_\pm = \alpha_0 \left(\frac{T}{T_{\rm c0}} -1 \mp \epsilon_0 \right)$. 
Thus, we obtain the parameter $\eta$ characterizing the chiral superconducting state, 
\begin{eqnarray}
\eta &=& \sqrt{\frac{\beta_1 \alpha_+ - (\beta_1 -2 \beta_2) \alpha_-}{\beta_1 \alpha_- - (\beta_1 -2 \beta_2) \alpha_+}}. 
\end{eqnarray}
Because $\alpha_\pm < 0$ and $|\alpha_-| < |\alpha_+|$, we obtain $\eta >1$ in the whole B-phase at $H=0$. 
When the temperature is decreased from $T_{\rm c}^{\rm B}$, the ratio of two-component order parameters changes from 
$\eta(T=T_{\rm c}^{\rm B}) = \infty$ to $\eta(T=0) >1 $. 
Thus, the superconducting state hosting unusual quadratic line node at $k_z=0$ is not realized at zero magnetic field. 

Now we turn to the superconducting state in the magnetic field. 
When the anisotropy in the Fermi surface is neglected, 
the mean field theory for the $f$-wave state gives a rather simple relations in the gradient term, 
$\kappa_3=\kappa_4=0$ and $\kappa_1 = \kappa_2$. 
Then, the gradient term is simplified as, 
\begin{eqnarray}
  F_0^{\rm grad}  &=& 
   \kappa_1 \left( |\vec{D} \eta_1 |^2 + |\vec{D} \eta_2 |^2 \right) 
  +\kappa_5 \left( |D_{\mathrm{z}}\eta_1 |^2 + |D_{\mathrm{z}}\eta_2 |^2 \right),  
  \label{eq:f1grad} 
\end{eqnarray}
with using $\vec{D} = \left(D_{\mathrm{x}}, D_{\mathrm{y}} \right)$. 
The symmetry breaking field gives rise to a correction to the gradient term 
\begin{eqnarray}
F_1^{\rm grad} &=& 
\epsilon_\parallel \left( |\vec{D} \eta_1 |^2 - |\vec{D} \eta_2 |^2 \right) 
  +\epsilon_{\perp} \left( |D_{\mathrm{z}}\eta_1 |^2 - |D_{\mathrm{z}}\eta_2 |^2 \right). 
\end{eqnarray}
The parameters are chosen as $\epsilon_\parallel <0$ and $\epsilon_\perp <0$ so as to reproduce 
the multiple phase diagram in UPt$_3$~\cite{Sauls}. 
Then, the orbital effect favors the $\Gamma_1$-state having a short coherence length. 
Thus, the $F_1^{\rm grad}$ term competes with the $F_1$ term. Since the former is more important 
than the latter at high magnetic fields, the $\Gamma_1$-state is stabilized in the high magnetic field C-phase. 
Thus, we obtain the parameter $\eta=0$ in the C-phase consistent with the list in Table~\ref{SM_tab2}. 
In the B-phase, the parameter $\eta$ is decreased through $\eta=1$ with increasing the magnetic field, 
and reaches to $\eta=0$ at the B-C-phase boundary. This scenario proposed by Sauls~\cite{Sauls} 
is consistent with the multiple superconducting phases in UPt$_3$ and has been regarded as the most promising 
scenario~\cite{Joynt}. We also adopt this scenario and assume the range of the parameter $\eta$ listed 
in Table~\ref{SM_tab2}.

\end{document}